\documentclass[aps,prb,preprint,groupedaddress,showpacs,amsfonts]{revtex4}
\usepackage{bm}
\usepackage{graphicx}

\begin{document}
% You should use BibTeX and revtex.bst for references
\bibliographystyle{apsrev}
% marks overfull lines with blackboxes
%\draft - no longer supported, use the 'draft' option instead

%\preprint{}

\title[FINITE-SIZE CALCULATIONS OF SPIN-LATTICE RELAXATION RATES IN $\ldots$]
{Finite-size calculations of spin-lattice relaxation rates in Heisenberg
spin-ladders}
% Optional argument for running titles on pages
%\title[]{}



\author{Martin P. Gelfand}
\email[]{gelfand@lamar.colostate.edu}
%\thanks{}
\author{Mohan Mahadevan}
\altaffiliation[Permanent address: ]%
{KLA-Tencor, 3 Technology Drive, Malpitas, California 95035}

\affiliation{Department of Physics, Colorado State University, Fort Collins, Colorado 80523}


\date{\today}

\begin{abstract}
Calculations of nuclear spin-lattice relaxation rates are carried
out by means of exact diagonalization on small ($2\times 6$)
antiferromagnetic Heisenberg ladders, using the simplest forms permitted
by symmetry
for the hyperfine couplings for the three nuclear sites in Cu$_2$O$_3$ ladders.
Several values of the rung/chain exchange ratio
$J_{\perp}/J_{\parallel}$ have been considered.
Comparisons with experimental results, field theoretic calculations,
and the Gaussian approximation highlight some open problems.
\end{abstract}

%\maketitle must follow title, authors, abstract and \pacs
\maketitle

% body of paper here - Use proper section commands (\section,\subsection)
\section{Introduction}
\label{sec:Introduction}
Spin-ladder systems, in particular the two-leg, $S=1/2$,
antiferromagnetic variety, have been the subject of considerable
theoretical and experimental investigation.\cite{dagotto96}
Spin ladders are appealing because they are one-dimensional
systems and thus can be effectively investigated using many 
powerful theoretical tools,
while offering a wider parameter space of ``simple," and potentially
experimentally realizable, Heisenberg Hamiltonians than spin chains.  
The simplest Heisenberg ladder Hamiltonian has the form
\begin{equation}
{\cal H}=\sum_n J_\parallel (\bm{S}_{n,1}\cdot\bm{S}_{n+1,1}
+ \bm{S}_{n,2}\cdot\bm{S}_{n+1,2}) + J_\perp \bm{S}_{n,1}\cdot\bm{S}_{n,2}
\label{eq:slham}
\end{equation}
which offers a dimensionless parameter $J_\perp/J_\parallel$ which is
in principle tunable by chemistry or pressure.
In addition, compounds containing weakly coupled
Cu$_2$O$_3$ ladders are appealing because of possible connections
with cuprate superconductivity.

The present work was motivated by the nuclear spin-lattice relaxation 
measurements in La$_6$Ca$_8$Cu$_{24}$O$_{41}$, an undoped ladder
compound, by Imai {\it et al.}\cite{imai98}  These measurements were
carried out for all of the nuclear sites on the ladder, namely the
copper, the ``rung'' oxygen, and the ``ladder'' (or ``chain'') oxygen,
over a wide temperature range, from low temperatures up to nearly
$900\,\rm K$.  Because the principal exchange interactions in cuprates
are so large, on the order of $1000\,\rm K$, it is quite challenging to 
do experimental work at temperatures significantly greater than the 
spin gap ($\Delta\approx 500\,\rm K$).

The experimental results (see Figure 1(c) of Imai {\it et al.}\cite{imai98})
have the following noteworthy features.
At temperatures below about $425\,\rm K$, the relaxation rates for all three
sites follow a common (activated) temperature dependence
up to a scale factor.
However, on increasing $T$ the copper $1/T_1$ (which we will refer
to as $1/\hbox{}^{\rm Cu}T_1$) exhibits a rather sharp
departure from that of the two oxygen sites ($1/\hbox{}^{\rm O(1)}T_1$
and $1/\hbox{}^{\rm O(2)}T_1$ for ladder and rung, respectively).  
There seems to be
a nearly discontinuous decrease in the derivative of $1/\hbox{}^{\rm Cu}T_1$;
moreover, above $425\,\rm K$ the $1/\hbox{}^{\rm Cu}T_1$ data appear nearly
linear with an almost vanishing intercept.
The relaxation rates for the two oxygen sites, in contrast,
exhibit no particular features in the vicinity of $425\,\rm K$.

Several aspects of the wavevector dependence of the low-frequency
spin susceptibility can be gleaned directly from the data.

One can express the spin-lattice relaxation rate in terms of
the dynamic structure factor for the Cu$^{2+}$ spins
\begin{equation}
{1\over\hbox{}^n T_1} \propto 
\int d\bm{q} H_n(\bm{q})S(\bm{q},\omega_n)
\end{equation}
where $H_n$ is the hyperfine form factor associated with nucleus $n$, 
$\omega_n$ is the NMR frequency (which we will take to be zero in
everything that follows), and $S$ is the structure factor.  The 
proportionality constants can be neglected for our present purposes.
The spin correlations are isotropic, so there is no need to consider 
the various components, $S^{xx}$ and so forth, individually.
The hyperfine interactions are not isotropic, so the orientation
of the magnetic field in the NMR experiment does affect the results;
however, all of the results of present interest can be obtained 
with a single field orientation, which then specifies $H_n(\bm{q})$ uniquely.
The largest hyperfine couplings are between a given nuclear site 
and the closest spins; at that level of approximation, and taking
the intra- and inter-chain lattice constants to be of unit length, one has
\begin{eqnarray}
H_{\rm Cu}=A^2,\ H_{\rm O(1)}=4C^2\cos^2(q_x/2),\nonumber\\
H_{\rm O(2)}=4F^2\cos^2(q_y/2)+D^2
\end{eqnarray}
where $C$, $F$, and $D$ are the hyperfine couplings identified
in Fig.~1(a) of Imai {\it et al.},\cite{imai98} $A$ is the on-site
hyperfine interaction for copper, and we have elided the 
orientation dependence of the hyperfine interactions
(so, for example, $A^2$ should really be $A_x^2 + A_y^2$ if the
static field is along the $z$ axis).

The essential difference between copper and oxygen sites is that
in the latter the hyperfine interaction in the vicinity of $\bm{q}=(\pi,\pi)$
is much smaller than in the vicinity of $\bm{q}=(0,0)$.
If, at all temperatures of experimental relevance, $S(\bm{q},0)$
had most of its weight in the vicinity of $\bm{q}=(0,0)$, then
then all three relaxation rates would have tracked one another.
The marked decrease of $1/\hbox{}^{\rm Cu}T_1$ relative to the other two
relaxation rates at $425\,\rm K$
indicates that this cannot be the case, and in fact suggests
that at temperatures below $425\,\rm K$ the ratio of the spectral weight near
$(\pi,\pi)$ to that near $(0,0)$ is roughly constant and of order unity, while
above $425\,\rm K$ the ratio falls markedly.  (The {\em decrease} is
crucial. If there were an increase in  $1/\hbox{}^{\rm Cu}T_1$ relative
to the oxygen rates with increasing $T$, one could ascribe
that to a turn-on of $S((\pi,\pi),0)$ for $T\agt\Delta$ but
$S((\pi,\pi),0)$ might have been negligible compared to 
$S((0,0),0)$ at lower temperatures.)

Why the emphasis on $\bm{q}=(0,0)$ and $(\pi,\pi)$?  In gapped systems
such as spin ladders, the low-energy spin fluctuations are Raman
processes, and at low temperatures one needs to consider only
the lowest energy magnons, namely those near $\bm{q}=(\pi,\pi)$.
Spin fluctuations near $(0,0)$ are associated with 
two-magnon processes, and those near $(\pi,\pi)$ with
three-magnon processes, and on the face of it one would
be justified in neglecting the three-magnon processes entirely
at low temperatures: see Ref.~\onlinecite{ivanov99} and
references cited therein.
However, as we have just seen, this appears to be inconsistent
with the experimental data for La$_6$Ca$_8$Cu$_{24}$O$_{41}$,
and it is also inconsistent with the quantum Monte Carlo
calculations of spin-lattice relaxation in a particular 
Heisenberg ladder ($J_\perp/J_\parallel=1$)
by Sandvik, Dagotto, and Scalapino,\cite{sandvik96}
at least at temperatures greater than half the magnon gap.

An extensive theoretical treatment of spin dynamics in
gapped one-dimensional Heisenberg models, including
spin ladders, has been presented by Damle and Sachdev.\cite{damle98}
Their analysis of $S(\bm{q},\omega\approx0)$ was restricted
to $\bm{q}$ near $(0,0)$, but they did find the quite interesting
result that the activation energy for $1/T_1$ is larger, by a
factor of $3/2$, than the activation energy for the uniform
static susceptibility (which is simply the spin gap).
An analysis of $S(\bm{q},\omega\approx0)$ for $\bm{q}$ near
$(\pi,\pi)$, for systems with $J_\parallel \gg J_\perp$  
has been presented by Ivanov and Lee.\cite{ivanov99}
Their results are suggestive of a fairly sharp crossover from
low- to high-temperature regimes at $T\approx\Delta$, and also
indicate that the $(\pi,\pi)$ contribution to $1/T_1$  ``overshoots"
its $T=\infty$ value and thus decreases as $T\to\infty$.

In the present work, we have applied
exact diagonalization to evaluate spin-lattice
relaxation rates, following the method of 
Sokol, Gagliano and Bacci.\cite{sokol93}  
We have considered three different ladder Hamiltonians, namely
$J_\perp/J_\parallel=0.5$, 1.0, and 2.0, and have obtained 
$1/T_1$ for Cu, O(1), and O(2) sites taking the simplest 
conceivable hyperfine couplings, namely $A=C=F=1$,
with all other interactions neglected.
All of the calculations were for rather small systems, $2\times6$,
such that exact diagonalization could be carried out in an
extremely straightforward manner.

It was noted above that calculations of spin lattice
relaxation rates for spin ladders have already been carried out
by means of large scale quantum Monte Carlo,\cite{sandvik96} but 
those calculations were limited to the Cu sites.  
Our goal is somewhat different than that of Sandvik, Dagotto, and
Scalapino's work.
We are not trying to fit the data in detail, rather 
we want to see what can
be learned from modest numerical calculations.  One reason 
not to fit the data is that to get the gap correct to
10\% by exact diagonalization for $J_\perp/J_\parallel=0.5$
would require a system at least $2\times12$.  Another is that
we do not treat the spin diffusion
contribution to the relaxation rates correctly: our
calculations effectively introduce an artificial cut-off so that
we obtain a finite spin-lattice relaxation rate.
Finally, the precise form of the
spin Hamiltonian for the cuprate ladder compounds is
still subject to argument.
Although the Knight-shift results of Imai {\it et al.}\cite{imai98}
appear to be consistent with the simple spin-ladder Hamiltonian
of Eq.~(\ref{eq:slham}) for $J_\perp/J_\parallel\approx0.5$,
it has been suggested by Brehmer {\it et al.}\cite{brehmer99} that
instead $J_\perp/J_\parallel\approx1$ and in addition 
there is a modest amount of plaquette ``ring exchange" in the
Hamiltonian.  A quantum-chemical analysis of the exchange
interactions in various cuprates\cite{mizonu98} provides some
support for the latter proposal, since it concludes
that $J_\perp/J_\parallel\approx0.9$ for Sr$_{14}$Cu$_{24}$O$_{41}$
(which is a lightly-self-doped version of the undoped
La$_6$Ca$_8$Cu$_{24}$O$_{41}$ compound).

To be precise, the goals of our calculation are as follows.
First, we want to verify that $S(\bm{q},0)$ has significant
weight near $\bm{q}=(\pi,\pi)$ as well as near $(0,0)$
and see if there are any noticeable trends with varying $J_\perp/J_\parallel$.
Second, we want to explore the crossover from low to high temperature
behavior in $1/T_1$: can we see anything like the experimental results,
or like the theoretical results of Ivanov and Lee?
Third, we want to keep our eyes open for any unanticipated patterns
that might emerge in the numerical results.

\section{Method of Calculation and Results}

The finite-size calculations of spin-lattice relaxation rates
are carried out following Sokol, Gagliano, and Bacci.\cite{sokol93}
Rather than repeating their discussion of the method let us make
a few remarks.  We take $J_\parallel$ as the unit of energy.

The first step in the calculation is a complete diagonalization
of the Hamiltonian and evaluation of matrix elements for certain
local spin operators (depending on which nuclear site one
is interested in).  For the $2\times 6$ lattices
all of the calculations could be done using the simplest possible
representations of the states in terms of local $S^z$ values; it
was not even necessary to use translational invariance to classify
states by wave vector.

The second step is the construction of an auxiliary function which
Sokol {\it et al.}\ refer to as $I(\omega)$. This is implicitly dependent
on $T$ and the hyperfine couplings.
We considered temperatures ranging from 0.3 to 50.
Typically we constructed $I(\omega)$ at intervals of 0.02 in $\omega$
up to at least $\omega=0.6$.

\begin{figure}
\includegraphics{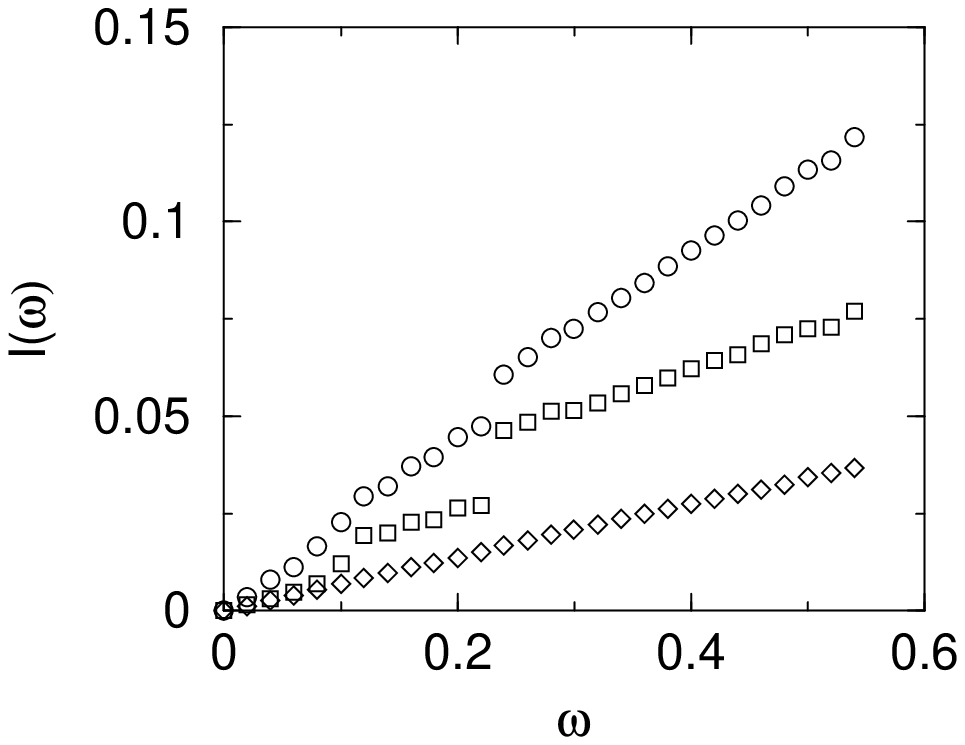}
\caption{The auxiliary function $I(\omega)$ for the Cu site with
$J_\perp=1.0$ and
$T=0.5$ (squares), 1.0 (circles), and 5.0 (diamonds).}
\label{fig:Iomega}
\end{figure}

\begin{figure}
\includegraphics{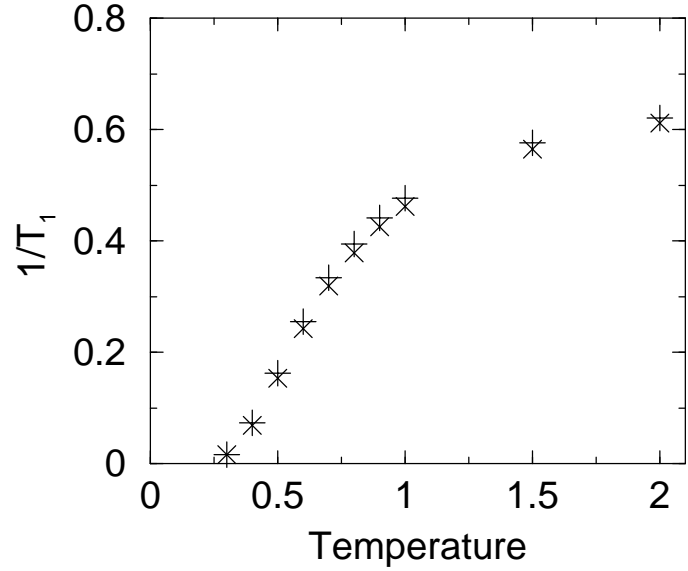}
\caption{Estimated values of $1/\hbox{}^{\rm Cu}T_1$ for 
$J_\perp=1.0$
as a function of temperature, taking $\omega_{\rm max}=0.3$ ($+$)
and 0.5 ($\times$).}
\label{fig:compcutoffs}
\end{figure}

Finally, one needs to estimate the zero-frequency derivative
of $I(\omega)$, because $1/T_1$ is is proportional to 
$T(\partial I/\partial\omega)|_{\omega=0}$.  At high temperatures
$I(\omega)$ is quite smooth, but at temperatures comparable to the
gap significant structure develops (see Fig.~\ref{fig:Iomega}).
In order to avoid introducing spurious temperature dependences
into $1/T_1$ it is important to use a consistent procedure for
extracting the derivative from the data.  What we did was to 
fit a zero-intercept line through all the data points up to a
cutoff $\omega_{\rm max}$, weighting all points equally in the fit.
We did all of the calculations using both $\omega_{\rm max}=0.5$ and
0.3.  While there are noticeable differences in the results using
these two cutoffs, as shown in Fig.~\ref{fig:compcutoffs}, our
conclusions turn out the same no matter which is chosen.  The use of a
much smaller cutoff, which might seem to be preferred on the grounds
that one is really looking for a zero-frequency derivative, is not
beneficial.  The structure that develops in $I(\omega)$ as $T$ is
lowered, making it look like a Devil's staircase, is
a finite-size artifact and must be averaged over, using a suitably
large $\omega_{\rm max}$, to obtain results that are 
representative of the thermodynamic limit.

%\begin{figure}
%\includegraphics[width=3.4in]{Iomega.eps}
%\caption{The auxiliary function $I(\omega)$ for the Cu site with
%$J_\perp=1.0$ and
%$T=0.5$ (squares), 1.0 (circles), and 5.0 (diamonds).}
%\label{fig:Iomega}
%\end{figure}

%\begin{figure}
%\includegraphics[width=3.4in]{compcutoffs.eps}
%\caption{Estimated values of $1/\hbox{}^{\rm Cu}T_1$ for 
%$J_\perp=1.0$
%as a function of temperature, taking $\omega_{\rm max}=0.3$ ($+$)
%and 0.5 ($\times$).}
%\label{fig:compcutoffs}
%\end{figure}

\begin{figure}
\includegraphics{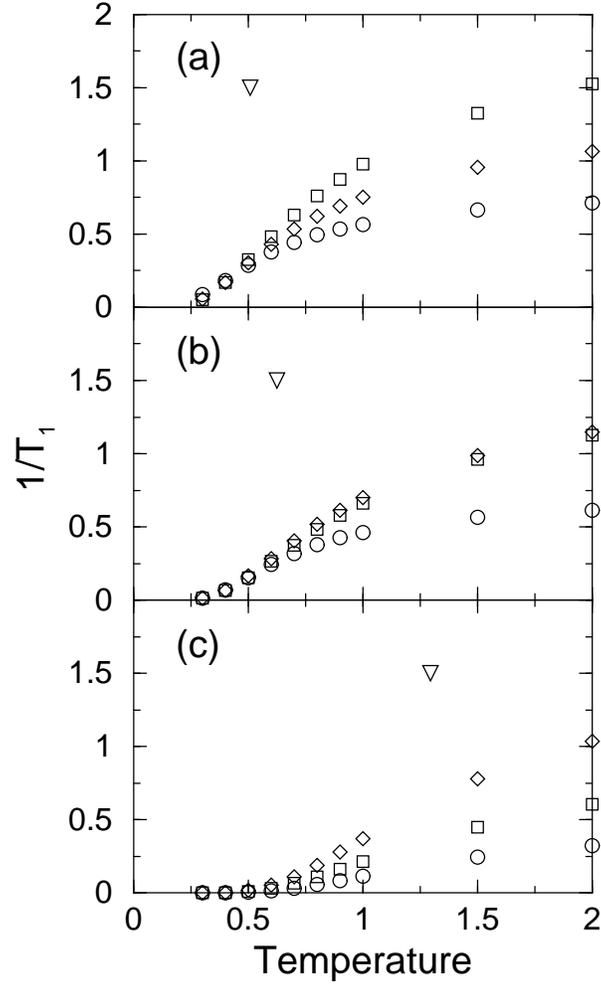}
\caption{Spin lattice relaxation rates as a function of temperature (in
units of $J_\parallel$)
for copper (circles), ladder oxygen (squares) and rung oxygen 
(diamonds) sites, for
$J_\perp/J_\parallel=0.5$, 1.0, and 2.0 in (a), (b), and (c) respectively.
The upside-down triangle on each graph indicates the value of
the spin gap $\Delta$ for the corresponding system.}
\label{fig:allT}
\end{figure}

We now turn to the results of the calculations for the three
nuclear sites and three values of $J_\perp$ considered (0.5, 1.0,
and 2.0).  In every case we take $\omega_{\rm max}=0.5$.
In Fig.~\ref{fig:allT} we present results on a linear 
temperature scale, for $T\leq2$.  The behavior of the spin-lattice
relaxation rate at high temperatures is a bit
surprising: comparing the plots in Fig.~\ref{fig:allT}(a) through
(c) it is apparent that while the Cu and O(1) rates decrease strongly
as $J_\perp$ increases, the trend for the O(2) rate is different.  
This is made more explicit in Fig.~\ref{fig:highT}, where we show
$1/T_1$ for all three sites as a function of $J_\perp$ at $T=50$
(effectively infinite temperature).  In contrast, at low temperatures
$1/T_1$ decreases with increasing $J_\perp$ at all sites, as
one would expect since the spin gap is an increasing function of $J_\perp$.

%\begin{figure}
%\includegraphics[width=3.4in]{plotall_2.eps}
%\caption{Spin lattice relaxation rates as a function of temperature (in
%units of $J_\parallel$)
%for copper (circles), ladder oxygen (squares) and rung oxygen 
%(diamonds) sites, for
%$J_\perp/J_\parallel=0.5$, 1.0, and 2.0 in (a), (b), and (c) respectively.
%The upside-down triangle on each graph indicates the value of
%the spin gap $\Delta$ for the corresponding system.}
%\label{fig:allT}
%\end{figure}

\section{Discussion and Conclusions}
\label{sec:conclusions}
It is evident that for $J_\perp=0.5$ and 1.0, 
$1/T_1$ for all three sites is nearly equal for temperatures
below the spin gap.  (Of course we do not claim that this holds
to arbitrarily low temperatures, just that it seems correct
for temperatures as low as we dare to estimate $1/T_1$.) 
Because of our choice of hyperfine interactions,
this suggests that in such cases the weight in $S(\bm{q},0)$ for
$\bm{q}\approx(\pi,\pi)$ is approximately three times that for 
$\bm{q}\approx(0,0)$.  This is in quantitative agreement with the results
of Sandvik {\it et al.}\cite{sandvik96} at $J_\perp=1.0$.
However, the story is rather different at $J_\perp=2.0$, where
the spin-lattice relaxation rates for all three sites, including
the two oxygen sites, are significantly
different even at $T=\Delta/2$.  In the strong-coupling limit,
then, the simple picture for $S(\bm{q},0)$ in which its weight
is concentrated at $(0,0)$ and $(\pi,\pi)$ does not work even
for temperatures that are a modest fraction of $\Delta$. 

What can we say about the low-to-high temperature crossover
in the spin-lattice relaxation rates?  First of all, the sort
of behavior seen experimentally, in which $1/T_1$ for the oxygen
sites track each other closely while $1/\hbox{}^{\rm Cu}T_1$ splits
off, appears to be a special feature of $J_\perp\approx1$ in
the present calculations; it is not at all generic and does
not hold for the putative experimental value $J_\perp\approx0.5$.  
Second, in no case does $1/\hbox{}^{\rm Cu}T_1$ exhibit any sort of
sharp ``break'' as seen experimentally; nor does $1/\hbox{}^{\rm Cu}T_1$
exhibit linear-in-$T$ behavior (with zero intercept, or otherwise)
in the high temperature regime, even over a restricted temperature
range (say $\Delta$ to $2\Delta$).  Finally, in no case does
$1/\hbox{}^{\rm Cu}T_1$ exhibit an ``overshoot'' during the crossover:
the spin-lattice relaxation rate associated with all sites
monotonically increases with $T$.

Our calculations thus suggest that there are quite a few open
problems in this field.  
Almost none of the prominent experimental facts concerning
$1/\hbox{}^{\rm Cu}T_1$ in La$_6$Ca$_8$Cu$_{24}$O$_{41}$
are reproduced in our finite-size calculations. 
Furthermore, the work of Ivanov and Lee\cite{ivanov99} does not
seem to have much to say about our results, either.  Their
calculation is controlled only in the $J_\perp\ll1$ regime, so
we should only look at the $J_\perp=0.5$ data. Here we have no
evidence of overshoot in $1/\hbox{}^{\rm Cu}T_1$, and no reason
to believe that one can just examine the spectral weight near
$(\pi,\pi)$ since  $1/\hbox{}^{\rm O(2)}T_1$ ``peels off'' from
$1/\hbox{}^{\rm O(1)}T_1$ in a manner not very different from
$1/\hbox{}^{\rm Cu}T_1$.

At this point we face several alternatives.  It is possible that
our results are simply unreliable, because we are
considering systems that are too small (especially for $J_\perp=0.5$)
and our procedure for estimating $dI(\omega)/d\omega$ is flawed.
We cannot rule this out, but we strongly suspect that the trends in the
results as a function of $J_\perp$ are robust.
It is possible that the 
spin Hamiltonian for the ladders in
La$_6$Ca$_8$Cu$_{24}$O$_{41}$ is more complicated than
the model we have considered.  Whether the Hamiltonian of
Brehmer {\it et al.}\cite{brehmer99} can reproduce the
spin-lattice relaxation data requires another calculation.
Another possibility that must be considered, given the
remarkably sharp feature in $1/\hbox{}^{\rm Cu}T_1$ found
in the experimental data, is that La$_6$Ca$_8$Cu$_{24}$O$_{41}$ undergoes,
by coincidence, a subtle structural transition at $425\,\rm K$.
This could introduce an anomalously strong $T$-dependence
to the hyperfine interactions, though why the effect should
be so much stronger in $H_{\rm Cu}(\bm{q})$ than 
$H_{\rm O(1)}(\bm{q})$ and $H_{\rm O(2)}(\bm{q})$
is difficult to envision.

\begin{figure}
\includegraphics{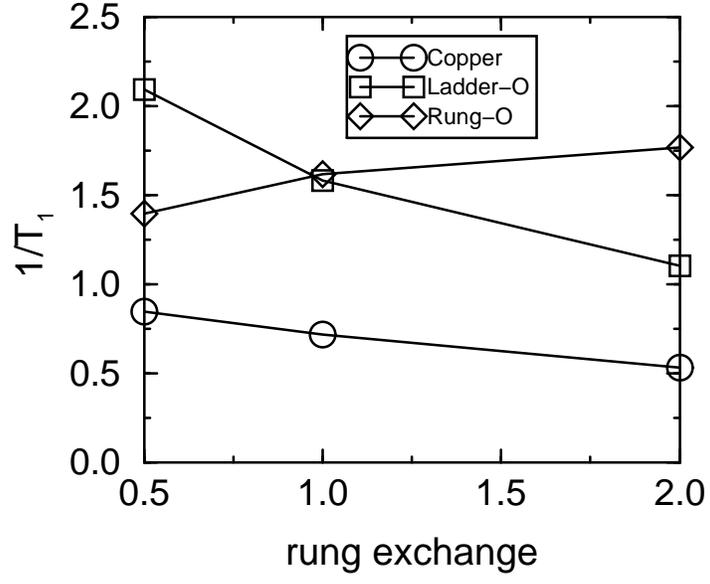}
\caption{Spin lattice relaxation rates at $T=50$ as a function of
$J_\perp$.}
\label{fig:highT}
\end{figure}

Let us now turn to the results of our calculations for spin-lattice 
relaxation at very high temperatures, shown in Fig.~\ref{fig:highT}.
The most natural way to think about these results is in terms of
the Gaussian approximation.\cite{anderson53,moriya56a,moriya56b}
The basic idea of this approach is to assume that 
$\int d\bm{q} H_n({\bm q}) S({\bm q},\omega)$ is a 
Gaussian function of $\omega$,
and then evaluate the frequency cumulants of this function
by means of short-time expansions of time-dependent correlation functions.
At $T=\infty$ the calculations are especially simple, because
the expectation values of correlators $\langle \bm{S}_i \cdot \bm{S}_j\rangle$
vanish for sites $i\neq j$.  For three sites of interest in
Heisenberg ladders, the Gaussian approximation yields the following
exchange dependences of the spin-lattice relaxation rates at $T=\infty$:
\begin{equation}
1/\hbox{}^{\rm Cu}T_1 \propto 1/\sqrt{1+\textstyle{1\over2}J_\perp^2}\ ,
\end{equation}
\begin{equation}
1/\hbox{}^{\rm O(1)}T_1 \propto 1/\sqrt{1+J_\perp^2}\ ,
\end{equation}
and $1/\hbox{}^{\rm O(2)}T_1$ does not have any $J_\perp$ dependence
at all.  (Recall that $J_\parallel \equiv 1$; in all of these
results there is an overall factor of $1/J_\parallel$.)
If this last result seems peculiar, let us note that it
can be derived in another way, by considering the strong-$J_\perp$
limit.  Then one most naturally thinks about the states in terms of
singlets and triplets on the rungs.  The relevant energy scale
for the dynamics of the total spin on a rung,
which is relevant to $1/\hbox{}^{\rm O(2)}T_1$, would seem to be
proportional to $J_\parallel$ (that is, the bandwidth in lowest-order
perturbation theory for a triplet
excitation in a single background\cite{barnes93}), and 
with the hypothesis of a single energy scale in
$\int d\bm{q} H_{\rm O(2)}S(\bm{q},\omega)$
one reproduces the Gaussian approximation result.

We see in Fig.~\ref{fig:highT} that $1/T_1$ for the copper and 
ladder oxygen sites decreases with increasing $J_\perp$, qualitatively
in agreement with the Gaussian approximation, although the dependence
on $J_\perp$ is not as strong as that approximation suggests.
Furthermore, $1/\hbox{}^{\rm O(2)}T_1$ exhibits an increase with
$J_\perp$.  The rather poor performance of the Gaussian approximation
is somewhat disappointing, considering how well it works for
estimating spin-lattice relaxation rates in square-lattice
Heisenberg antiferromagnets.\cite{gelfand93,sokol93}  It is not too surprising,
perhaps, given that the dynamic correlations in the $S=1/2$ Heisenberg chain
are far from Gaussian at $T=\infty$.\cite{roldan86}
So, there is yet another open problem in the area of low-energy spin dynamics
of Heisenberg ladders.

%\begin{figure}
%\includegraphics[width=3.4in]{highT.eps}
%\caption{Spin lattice relaxation rates at $T=50$ as a function of
%$J_\perp$.}
%\label{fig:highT}
%\end{figure}

% If in twocolumn mode, this environment will move to single column
% format so that long equations can be displayed. Use
% sparingly. Requires multicol.sty (automatically read in by the
% twocolumn option).
%\begin{widetext}
%\end{widetext}

% If you have acknowledgements this puts in the proper section
\acknowledgements
This work was supported by the US National Science Foundation
through grant DMR 94--57928.  We thank T. Imai for several stimulating
discussions and also for communicating the results of his group's
experiments prior to publication.

% Create the reference section using BibTeX
\bibliography{spinladder}

% figures follow here or may be put into the text as floats.
% Use the graphics or
% graphicx packages distributed with LaTeX2e. See the LaTeX Graphics
% Companion by Michel Goosens, Sebastian Rahtz, and Frank Mittelbach
% for instance.

% tables follow here or maybe be put in the text
% \begin{table}
% \caption{}
% \label{}
% \end{tabular}
% \end{table}

\end{document}